%% file: elsarticle-MAIN.tex
\documentclass[12pt]{elsarticle}

\usepackage{amssymb}
\usepackage{multirow}
\usepackage[table,xcdraw]{xcolor}
\usepackage{graphicx}
\usepackage{pdflscape}
\usepackage{booktabs}
\usepackage{threeparttable}
\usepackage{longtable} % for 'longtable' environment
\usepackage{hyperref}
\journal{Network and Computer Applications}

\begin{document}

\begin{frontmatter}

%% Title, authors and addresses

%% use the tnoteref command within \title for footnotes;
%% use the tnotetext command for theassociated footnote;
%% use the fnref command within \author or \address for footnotes;
%% use the fntext command for theassociated footnote;
%% use the corref command within \author for corresponding author footnotes;
%% use the cortext command for theassociated footnote;
%% use the ead command for the email address,
%% and the form \ead[url] for the home page:
%% \title{Title\tnoteref{label1}}
%% \tnotetext[label1]{}
%% \author{Name\corref{cor1}\fnref{label2}}
%% \ead{email address}
%% \ead[url]{home page}
%% \fntext[label2]{}
%% \cortext[cor1]{}
%% \affiliation{organization={},
%%             addressline={},
%%             city={},
%%             postcode={},
%%             state={},
%%             country={}}
%% \fntext[label3]{}

\title{Systematic review and characterisation of malicious industrial network traffic datasets} %for aiding ML algorithm performance testing}

%% use optional labels to link authors explicitly to addresses:
%% \author[label1,label2]{}
%% \affiliation[label1]{organization={},
%%             addressline={},
%%             city={},
%%             postcode={},
%%             state={},
%%             country={}}
%%
%% \affiliation[label2]{organization={},
%%             addressline={},
%%             city={},
%%             postcode={},
%%             state={},
%%             country={}}

\author[inst1]{Martin Dobler}

\affiliation[inst1]{organization={Research Centre Business Informatics, Centre for Robust Decision Making, Vorarlberg Univeristy of Applied Sciences},%Department and Organization
            addressline={Campus V, Hochschulstr. 1}, 
            city={Dornbirn},
            country={Austria}}

\author[inst1]{Michael Hellwig}
\author[inst3,inst4]{Nuno Lopes}
\author[inst5]{Ken Oakley}
\author[inst5]{Mike Winterburn}

% \affiliation[inst2]{organization={JR-Centre for Robust Decision Making, Vorarlberg Univeristy of Applied Sciences}%Department and Organization
            % }
\affiliation[inst3]{organization={2Ai – School of Technology, Polytechnic Institute of Cávado and Ave},
            city={Barcelos},
            country={Portugal}}

\affiliation[inst4]{organization={LASI – Associate Laboratory of Intelligent Systems},
            city={Guimarães},
            country={Portugal},
             }

\affiliation[inst5]{organization={Technological University of the Shannon},
            addressline={Moylish Park}, 
            city={Limerick},
            country={Ireland}}

\begin{abstract}
%% Text of abstract
The adoption of the Industrial Internet of Things (IIoT) as a complementary technology to Operational Technology (OT) has enabled a new level of standardised data access and process visibility. This convergence of Information Technology (IT), OT, and IIoT has also created new cybersecurity vulnerabilities and risks that must be managed. Artificial Intelligence (AI) is emerging as a powerful tool to monitor OT/IIoT networks for malicious activity and is a highly active area of research. AI researchers are applying advanced Machine Learning (ML) and Deep Learning (DL) techniques to the detection of anomalous or malicious activity in network traffic. They typically use datasets derived from IoT/IIoT/OT network traffic captures to measure the performance of their proposed approaches. Therefore, there is a widespread need for datasets for algorithm testing. This work systematically reviews publicly available network traffic capture-based datasets, including categorisation of contained attack types, review of metadata, and statistical as well as complexity analysis. Each dataset is analysed to provide researchers with metadata that can be used to select the best dataset for their research question. This results in an added benefit to the community as researchers can select datasets more easily and according to specific Machine Learning goals. 
\end{abstract}

%%Research highlights
\begin{highlights}
\item Industrial networks provide data access but pose cybersecurity risks 
\item Artificial Intelligence is used as a tool to support existing cybersecurity approaches
\item Machine Learning is used to detect anomalous/malicious industrial network traffic
\item Numerous public industrial test datasets for researchers and practitioners exist
\item Publicly available industrial datasets are systematically reviewed and compared
\item Datasets are analysed to determine attack types, statistics, and complexity scores

\end{highlights}

\begin{keyword}
%% keywords here, in the form: keyword \sep keyword
Machine Learning \sep Industrial Cybersecurity \sep Operational Technology (OT) \sep Industrial Internet of Things (IIoT) \sep Malicious Network Traffic \sep Dataset Complexity 
%% PACS codes here, in the form: \PACS code \sep code
\PACS 0000 \sep 1111
%% MSC codes here, in the form: \MSC code \sep code
%% or \MSC[2008] code \sep code (2000 is the default)
\MSC 0000 \sep 1111
\end{keyword}

\end{frontmatter}

%% \linenumbers

%% main text
\section{Introduction}
\label{sec:intro}

Industry 4.0 is driving the convergence of IT and OT and is resulting in the wider deployment of IIoT technologies, giving rise to complex heterogeneous distributed systems, particularly in industrial and manufacturing environments \cite{NIST2023}. Such systems can consist of i) OT such as Programmable Logic Controllers (PLC), computer numerical control systems, and complex Industrial Control Systems (ICS) including the supervisory control and data acquisition systems, ii) IT such as databases, data integration tools, data visualisation systems, reporting systems, AI systems and additionally iii) IIoT such as distributed sensors, smart robotics, track and trace systems and even augmented reality \cite{NIST2018}.

Consequently, Industry 4.0 cybersecurity management has become a difficult challenge since systems communicate using a wide variety of standards and protocols, sometimes distributed over large geographic areas. They use a diverse mix of hardware and software components and consist of varied physical infrastructure, meaning that traditional cybersecurity management tools such as firewalls, Intrusion Detection Systems and Intrusion Prevention Systems (IDS/IPS), application log monitors, and network monitoring tools are becoming increasingly limited in their ability to contain cybersecurity risk \cite{NIST2018,Toth2022}.

Researchers, practitioners as well as developers of said cybersecurity tools are seeking ways to use Artificial Intelligence (AI), especially Machine Learning (ML) or Deep Learning (DL), to automate intrusion detection. To facilitate the training of such systems they require test and/or production network traffic and data to test their algorithms, toolchains, and overall approaches. IT networks traditionally used for testing IDS/IPS such as campus LANS and IoT test bed environments have received significant attention, but research into OT and IIoT environments is lacking\cite{Alruwaili2021}. 
In this paper, we seek to identify datasets that contain malicious traffic, or anomalous data in industrial environments, for analysts to train AI/ML/DL intrusion detection systems. For these environments, IT generally refers to traditional Local Area Networks (LAN) and Internet traffic using Transmission Control Protocol/Internet Protocol (TCP/IP), while OT may include IoT, IIoT and ICS traffic~\cite{Stouffer2023}. IoT refers to the global network of physical devices or ‘things’ that use TCP/IP and the Internet to connect and exchange data~\cite{Booij2022}. IIoT is a subset of IoT and has smart sensors and actuators used in the industrial process, also using TCP/IP, and may be linked with Edge computing~\cite{NIST2018}. ICS is composed of industrial control systems such as SCADA, DCS, PLCs, and BAS, among others, using non-TCP/IP protocols but may additionally use a variety of traditional TCP/IP protocols, as presented in Table~\ref{tab:protocols}~\cite{SANS2022}.
\begin{table}[t]
\centering\renewcommand{\arraystretch}{1.}
\caption{Network protocols used in industrial environments.}
\begin{tabular}{|l||p{3.5cm}|p{3.5cm}|}\hline
\textbf{Common IT Protocols} & \multicolumn{2}{c|}{\textbf{Common ICS/OT Protocols}}   \\\hline
DHCP     & BACnet      & IEC104           \\
DNS          & BGAN        & Modbus TCP            \\
IEEE 802.11         & DHCP        & OPC                \\
HTTP    & DNP3        & PROFIBUS                  \\
HTTPS   & DNS         & PROFINET            \\
SFTP     & Ethernet/IP & SMB                 \\
SMB    & FTP         & SSH            \\
SMTP      & HART        & Telnet              \\
SSH         & HTTP        & VSAT   \\
HTTP       & HTTPS       &               \\
  & ICCP        &   \small{ + various industrial}    \\
     & IEC101      &   \small{ \quad proprietary ones} \\ \hline
\end{tabular}
\label{tab:protocols}
\end{table}

The main objective of this work is to provide the Industrial cyber security research community with a single source of information covering modern malicious IIoT and OT network traffic datasets. This enables the community to more easily select the best dataset to use for their own OT intrusion detection research. The contributions to research and sub-objectives are:
\begin{itemize}
    \item to identify publicly available IT, IoT, IIoT, OT and ICS malicious network traffic datasets;
    \item to carry out a descriptive analysis of these datasets;
    \item to identify the OT datasets that are most relevant to OT cyber security research;
    \item to identify the attack types in these OT datasets; and
    \item to provide metadata, descriptive statistics, as well as complexity analysis on these OT datasets.
\end{itemize}

The remainder of this work is structured as follows. In section~\ref{sec:relatedwork}, we discuss related work in scientific publications, examine existing definitions of the term OT as well as analyse previous meta-analysis of comparable datasets. Section~\ref{sec:method} describes methods used for researching datasets, dataset identification as well as dataset analysis, especially high-level extraction of relevant meta-data, and specific analysis of relevant features. Section~\ref{sec:results} presents results on the dataset analysis regarding the attack categorisation, complexity analysis as well as feature analysis. The results are then discussed in section~\ref{sec:discussion}. 
Section~\ref{sec:conclusion} concludes this work and provides recommendations for practitioners and researchers as well provides directions for future work. 

%%%%%%%%%%%%%%%%%%%%% NEXT SECTION %%%%%%%%%%%%%
%%%
%%%

\section{Related Work}
\label{sec:relatedwork}

Related work is often based on analysis of ML algorithm performance, adequacy of datasets or IDS-related use cases which might include more specific technological distinctions such as honey pot systems, honey net systems, network monitoring, or detection/prediction of unknown attack vectors. 

Meta-analysis and taxonomies for attack types and vectors do exist in IT security research, albeit without directly linking them to dataset or network capture analysis. More commonly, attack categorisation is done in conjunction with analysis of ML algorithm performance for IDS. Yi et al. \cite{Yi2023}, for example, give a five-generation overview of cyber-attacks, ranging from virus attacks in the 1980s to application attacks in the 2000s in which attackers target specific vulnerabilities to modern fifth-generation large-scale, multi-dimensional cyber-attacks (‘Mega attacks’). They argue that there is a relationship between attack and attack detection technologies, named ‘spears’ and ‘shields’, where intrusion detection based on machine learning is the latest and most modern ‘shield’ deployed against large-scale ‘weapons-grade’ attacks. In addition, they give an overview of common machine learning algorithms and argue that more research is needed on model stability and incremental learning, i.e., real-time identification of cyber-attacks based on online learning, for which they provide an analysis of 21 applications of machine learning and compare their robustness and accuracy.

Similarly, ten open-access datasets are analysed by Ahmetoglu \& Das \cite{Ahmetoglu2022} using common machine learning metrics and algorithms. They give classes of ML algorithms: shallow machine learning methods, supervised deep learning methods and unsupervised, or semi-supervised, ML/DL methods. Their comparison leads to the conclusion that dimension reduction and feature selection methods are critical for an IDS’s performance and constantly updated datasets with an adequate attack variety in them enable the development of (semi-) automated intrusion detection systems. 

Nisha \& Pramod \cite{Nisha2023}, focus on anomaly-based intrusion detection methods specifically designed for detecting insider attacks. The study examines host-based and network-based anomaly detection techniques, thus exploring the concepts of event-based intrusion detection by leveraging characteristics of sequential network data. The research identifies the potential of these techniques and suggests combining the strengths of anomaly-based, signature-based, and knowledge-based models to proactively detect attacks. 

In a survey of Federated Learning in IDS solutions, Agrawal et al. \cite{Agrawal2022} focus on security, privacy, and reliability. They address the challenges and vulnerabilities associated with such implementations, including high latency, false alarms, and poisoning attacks, and highlight their negative impact on different aspects of IDS. Furthermore, the paper explores future research opportunities and suggests potential solutions to address the challenges faced in implementing Federated Learning in IDS, including a list of future directions which IDS developers must take to incorporate AI techniques. As a derived implication, dataset providers must also consider providing relevant test use cases and scenarios. 

Ahmad et al. \cite{Ahmad2021}, reviewed network intrusion detection mechanisms based on ML and deep learning methods. They provide classification schemes, and methodologies used in various articles, highlighting a recent trend towards Deep Learning-based methodologies, particularly using autoencoders (AE) and deep neural networks (DNN). AE and DNN exhibit superior performance in terms of accuracy and false alarm reduction. However, the complexity and resource requirements of DL models pose challenges for real-time network intrusion detection. The paper emphasises the need for testing models with updated datasets to address modern network attacks and identifies research gaps in improving model performance for low-frequency attacks and in reducing complexity. In addition, the study reveals that 60\% of the methodologies examined in the research utilised the KDD Cup'99\footnote{\url{ http://kdd.ics.uci.edu/databases/kddcup99/kddcup99.html}} and NSL-KDD \cite{Tavallaee2009} datasets due to their results, availability, ease-of-use, as well as their status as de-facto standards. However, these datasets, dating from before 2010, might be considered outdated and do not adequately address modern network attacks, thereby limiting the performance in real-time environments. The authors suggest enhancing detection accuracy for intrusions in NIDS by using models with more up-to-date datasets.

Moreover, we observe a need to establish a clear differentiation between datasets and the technological context or scenarios for their intended ML use. Scenarios can include IoT networks, IIoT applications, ICS and OT. While IoT generally refers to interconnected devices and systems, IIoT specifically pertains to industrial applications, and ICS and OT encompass the practice of using hardware and software to control industrial equipment. Notably, OT -- and to a lesser extent ICS -- interface directly with the physical world. 

It is worth noting that numerous individual datasets and related scientific publications exist for the above scenarios \cite{Conti2021}. However, the availability of meta-analyses, such as taxonomies or comparative analyses of datasets, is relatively scarce compared to the plethora of available datasets. Also, most publicly available datasets predominantly focus on IP data, while industrial data, particularly from Supervisory Control and Data Acquisition (SCADA) networks or similar, is relatively uncommon.

At this point, we need to look at the varying definitions for the term OT that exist in literature. In particular, the differentiation between IIoT, IoT, ICS, and OT is not always easy to make, especially when considering whether IIoT and IoT in industrial applications are subsets of OT or OT is a separate term altogether. For this paper, OT consists of programmable systems and electro-mechanical devices which interact with a physical environment in industrial settings \cite{Hahn2016}. The definition is based upon the National Institute of Standards and Technology (NIST) approach towards security in OT as described in their Guide to Operational Technology (OT) Security \cite{Stouffer2023}. Technologies and terminologies included as subsets of OT include SCADA, “\emph{distributed control systems (DCS), programmable logic controllers (PLCs), building automation systems (BAS), physical access control systems (PACS), and the Industrial Internet of Things (IIoT).}”

Work relating to application scenario categorisation of datasets was done by Chatterjee \& Ahmed \cite{Chatterjee2022} in their review of 64 recent publications and derived application areas of anomaly detection for IoT devices. They list specific application areas e.g. Health Machinery, Robotics or Smart Cities. They also argue that modern datasets must address four issues: 
\begin{enumerate}
    \item data imbalance;
    \item production data pipelines can have a multitude of data sources which algorithms are not able to handle at scale;
    \item data drift requires frequent re-training of models;
    \item methods for reliably generating augmented data are required for applications with insufficient data.
\end{enumerate}
For the technological categorisation of Network Intrusion Detection System (NIDS) solutions, Gyamfi \& Arcut \cite{Gyamfi2022} provide a taxonomy and tabular classification of detection methods, NIDS placement strategies, security threats, and validation methods.

\section{Methodology}
\label{sec:method}

\begin{figure}[htbp]
\centering
\includegraphics[width=0.98\textwidth]{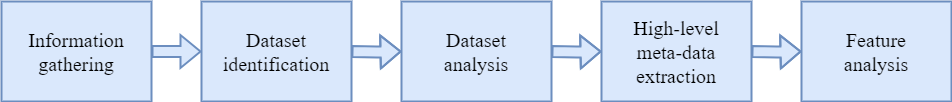}
\caption{Overall methodology workflow}
\label{fig:methodology}
\end{figure}

In this section, we present the methods that are used for information gathering, dataset identification, dataset analysis which included high-level extraction of relevant meta-data, and specific analysis of relevant features, as depicted in Figure~\ref{fig:methodology}. Details are provided in the respective subsections.

\subsection{Information Gathering Method}
\label{sec:31}
A systematic literature review was conducted to help identify different datasets relevant to ML training and testing IT, IoT, IIoT, ICS and OT network intrusion detection systems. Scopus and Google Scholar were used as the primary source of research papers, as well as papers referenced within the discovered papers. The search terms used are shown in Table~\ref{tab:searchterms}:

\begin{table}[htbp]\centering
\caption{Search Terms of literature review.}
\renewcommand{\arraystretch}{1.1}
\begin{tabular}{|c p{8.5cm}|}\hline
\textbf{No.} & \textbf{Direct dataset search terms}   \\ \hline
 1. &	IoT network anomaly detection dataset \\
2.	&IIoT network anomaly detection dataset\\
3.&	OT network anomaly detection dataset\\
4.&	ICS network anomaly detection dataset\\
5.&	Scada network anomaly detection dataset\\
6.	&Smart factory anomaly detection dataset\\
7.&	Industry 4.0 network anomaly detection dataset\\
8.&	Network Intrusion detection dataset\\\hline
\end{tabular}
\label{tab:searchterms}
\end{table}

\subsection{Dataset Identification}
\label{sec:32}
\begin{figure}[b]
\centering
\includegraphics[width=0.65\textwidth]{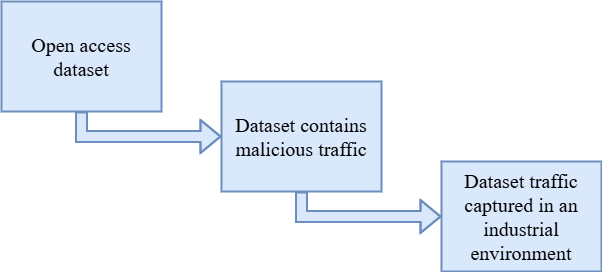}
\caption{Dataset Filtering Workflow}
\label{fig:filtering}
\end{figure}

The search mechanism applied an AND operator to the individual words in the search term. We carried out the search on 1st July 2023, filtered results to include the 5 years from 2019 to 2023, and initially sorted the results by 'Cited by (highest)', i.e., the number of citations and then as a second sort by 'Date (newest)' in case newer papers had not been referenced. The first 25 results for each search term and sorting approach were selected and the papers reviewed for references to datasets. As a result a total of 94 datasets with potential interest in IoT, IIoT, and OT intrusion detection research were discovered. We further filtered the identified datasets by the criteria illustrated in Figure~\ref{fig:filtering}.

This allowed us to identify 32 open-access industrial traffic datasets that contained malicious traffic for further analysis and comply with the following seven criteria:
\begin{enumerate}
    \item The result references a dataset that can be downloaded from a single website URL with the creator's copyright permission (possibly requesting access but with no charge).
    \item The dataset has been used in scholarly work as evidenced by a Scopus or Google Scholar search for the dataset name.
    \item The dataset contains both industrial and malicious traffic.
    \item Any IoT dataset was from an industrial environment and not only domestic appliances.
    \item The data was sufficiently well documented to be able to understand it.
    \item The data was labelled as benign or malicious.
    \item The related article was cited a minimum of 5 times.
\end{enumerate}

\subsection{Cyber-Attack Framework Selection and Attack Mapping}
\label{sec:33}
To further refine the relevance of the 32 datasets we reviewed the documentation that was available for each dataset and identified a total of 97 different cyber-attacks.

There may be an amount of overlap amongst attacks as different authors may record attacks slightly differently, e.g., DoS and DDoS etc. There are two commonly referenced cyber-attack frameworks, namely the Cyber Kill Chain (CKC) \cite{LockheedMartin2011} and the Mitre ATT\&CK framework \cite{Strom2020}. As the main intent of this paper is to help researchers identify datasets for further research, we chose the CKC and not the Mitre ATT\&CK framework. CKC provides an easily comprehensible approach through a seven-step chain, where each step is an attack vector for a threat actor. Whereas the Mitre ATT\&CK framework provides greater granularity but arguably more detail than needed for researchers to identify suitable malicious traffic datasets for research projects. Resultantly, we mapped the 97 identified cyber-attacks to the CKC framework.

\subsection{Exploratory Analysis}
Exploratory analysis was carried out on the 32 datasets.  The objective was to identify dataset files for deeper analysis.  We used the concept of “ML-ready” to select appropriate files.  ML-ready refers to a dataset file that has been pre-processed and structured in a way that it can be directly used as input for machine learning models.  The raw network data must have been transformed into a clean, consistent format with labels.  Where possible, we loaded dataset files into a python environment and reviewed the number of features, type of labels, and number of rows.  Just 17 of the 32 datasets had files that could be loaded in this way. The other datasets were not ML ready and hence were discounted.  

\subsection{Detailed Analysis}
\label{sec:35}
After identifying the IIoT/OT datasets we began to consider metrics that could be used to aid ML/AI practitioners when selecting datasets for research.  One such metric is complexity.

The complexity of test datasets is important to consider when testing ML algorithms. In data science, the term complexity refers to the variety and distribution of the data points contained in a dataset \cite{Barella2021, Garcia2018}. Complex datasets are characterised by numerous factors, including the size of the dataset, the dimensionality of the features, and the presence of outliers or missing values. Complexity can also result from the variety of data types, e.g., numeric, categorical, or textual data and the relationships between data points. A complex dataset often poses a greater challenge for pre-processing, modelling, and interpretation as it requires more sophisticated techniques to process. Therefore, understanding the complexity of datasets is essential for selecting and evaluating ML models. In the context of datasets for classification problems, such as the detection of attacks in OT network traffic, features such as the imbalance ratio or the class overlap also affect the complexity rating.

Various metrics and measures do exist to estimate the complexity of datasets \cite{Lorena2020}. These help to gain a deeper understanding of the data. Common complexity metrics include the number of features (dimensionality), data distribution characteristics, entropy, and information gain. In addition, measures such as the Gini index and Shannon entropy can be used to assess the diversity of categorical data. Methods for detecting outliers, such as the interquartile range or Mahalanobis distance, help identify data points that deviate significantly from the norm. Metrics such as sparsity of data and imbalance ratio are critical for datasets with class imbalances.

The metrics can be roughly categorised into six classes: 
\begin{enumerate}
    \item feature-based measures,
    \item linearity measures,
    \item neighbourhood measures,
    \item network measures,
    \item dimensionality measures,
    \item class imbalance measures.
\end{enumerate} 
While feature-based measures characterise how informative the available features are to separate the classes, linearity measures try to quantify whether the classes can be linearly separated, e.g., by calculating the distance of incorrectly classified samples from a linear Support Vector Machine (SVM) hyperplane. Neighbourhood measures characterize the presence and density of the same or different classes in the proximity of a data point. Examples are the R-value metric which measures the categorical class overlap \cite{Oh2011} or the hostility measure \cite{Lancho2023}. By conceiving the dataset as a mathematical graph, network measures try to extract structural information from graph features, e.g., the density of a k-Nearest Neighbours graph. Dimensionality measures evaluate data sparsity based on the number of samples relative to the data dimensionality and class imbalance measures consider the ratio of the number of examples between classes \cite{Goyal2021}. 

\begin{figure}[t]
\centering
\includegraphics[width=0.95\textwidth]{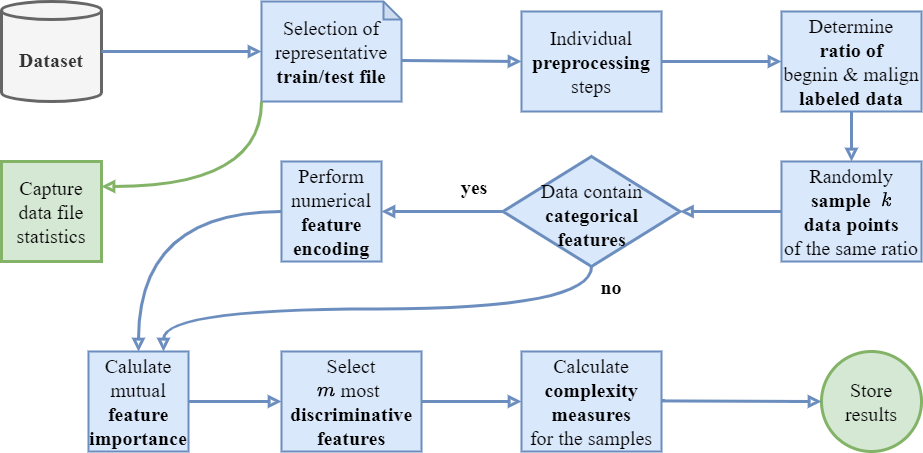}
\caption{Workflow of the dataset analysis leading to Table~\ref{tab:cs-results}.}
\label{fig:analysisworkflow}
\end{figure}

A recently developed tool for the direct calculation of a widely recognised subset of these complexity measures is the Problexity module \cite{Komorniczak2023}. The Problexity module represents an open-source easy-to-use Python library containing the implementation of 22 measures that assess individual aspects of the complexity of datasets for classification and regression problems. The package supports the calculation, analysis and visualisation of its problem complexity measures.
The Problexity module is used below (together with other key figures) to give an initial impression of the complexity of a dataset for ML purposes. Due to the size of individual datasets, these calculations could not always be carried out for the entire dataset due to a lack of computational resources. An analysis workflow is therefore proposed that approximates the dataset based on a representative sample. They are therefore only intended to provide an initial impression for the user and facilitate the selection of a dataset for individual purposes. Figure~\ref{fig:analysisworkflow} illustrates the major steps for the preparation, understanding, and average classification complexity analysis of the attack datasets.

The datasets analysed vary greatly in their structure and the quality of the data preparation. In addition, the datasets may contain a multitude of individual files with information on use-case specific raw network traffic data formats or files with unlabelled data. Some sets include files already prepared for ML applications. As analysing all raw data and processing network traffic for each dataset requires a lot of computational resources and network capture automation goes beyond the scope of the paper, the first step is to select a file that is as representative as possible, labelled, and suitable for ML applications. This file then forms the basis for the subsequent analyses. The characteristic key figures such as file size, the number of features, the number of data points, or the imbalance ratio (IR) are collected for the selected file.

Individual pre-processing steps are then carried out to clean the data and prepare the following analysis steps. This step involves dropping data points with missing feature values as well as unification of label data types. Further, data points with different attack flags are aggregated under the label ‘malicious’ to create a binary classification problem. Calculating the complexity value for large numbers of data points can be time-consuming. Hence, the complexity value is approximated by the use of a representative sample of all data points. For this purpose, it is necessary to determine the ratio of data points labelled as benign and malicious. Afterwards, $k$ data points are randomly sampled ensuring the very same ratio. 

The next step determines whether the dataset contains categorical features, which in this case must be represented by numerical values for the following calculations using one-hot encoding \cite{Pedregosa2011}. If there are no categorical features, the feature encoding step is skipped. 

Feature importance measures can help you identify which features are most relevant to the problem of classifying the data with respect to the binary labels. Based on the feature importance, a subset of the $m$ most discriminative features is selected to reduce dimensionality. Making use of the $\textsc{mutual\_info\_classif}$ function from the Scikit-Learn $\textsc{feature\_selection}$ library \cite{Pedregosa2011}, the Mutual Information (MI) values between each feature and the label for each dataset are calculated. The $m$ features with the highest mutual information are then selected from each dataset. Mutual information was chosen because all the datasets had mixed numerical and categorical features as well as categorical labels. Also, mutual information can reveal nonlinear relationships in the data that may not be identified by other filtering methods. 

Consequently, the average complexity score (CS) of the $k$ data points is calculated based on these m features. The complexity score indicates the variability within the data and may influence the choice of ML algorithms.

In addition to the complexity analysis of the suitable datasets, a more detailed investigation is carried out on a small subset of the OT datasets to better understand the link between features and complexity score. This approach enables us to take a closer look at the composition of the data sets as well as to draw insights into the complexity scores received. To this end, exemplary datasets with high, medium, and low complexity scores are selected from the ML ready datasets. In order not to go beyond the scope of this review, we will limit ourselves to three specimen data sets.

Analogously to the previous analysis steps, the three selected datasets are filtered based on mutual information to select the m feature columns with the highest shared information between the features and the target labels. To conceive a univariate visualisation of the class boundaries and overlap for each feature, individual feature plots are created for each of the three datasets. For each of the m features of a dataset, the univariate distribution of feature values is plotted against the row index of data points for all the rows in the selected file. The horizontal axis of each plot is the row index in the analysed CSV file and the vertical axis is the feature value on a linear scale from the feature’s minimum value to its maximum value. Additionally, colour encoding is used to distinguish benign (blue) and malicious (red) traffic inside these feature plots.  

%%%%%%%%%%%% NEXT SECTION %%%%%%%%%%%%%%%
%%%
%%%

\section{Results}
\label{sec:results}

In this section, we present our results on the dataset analysis regarding the attack categorisation, complexity analysis and feature analysis.
The 32 open-access industrial traffic datasets selected in the literature review process, section~\ref{sec:method}, are presented in Table~\ref{tab:32datasets}. For each dataset, the year of origin, the specific field of application from which the data is obtained, as well as a reference of a corresponding research paper (or web source in those cases where no paper could be referenced) are shown.
The datasets were originally published from 2011 to 2023 with the majority dating from 2018 or later.  We identified and classified the dataset application scenarios into 5 high level classifications: Critical Infrastructure, Industrial Systems, Military, IoT and IIoT, with more granular sub classifications also shown.
The following sections take a closer look at different aspects of these datasets, or subsets of the datasets, respectively. In Section~\ref{sec:41}, we examine the number and type of different attacks within the data. Details on the classification complexity of the ML-ready datasets are provided in Section~\ref{sec:42}. Finally, Section~\ref{sec:43} focuses on the feature analysis of three exemplary datasets of varying complexity.

\subsection{Attack Categorisation}
\label{sec:41}
Xenofontos et al. \cite{Xenofontos2022} survey multiple attack taxonomies for IoT and propose multiple attack types present in IoT networks. All these taxonomies follow a similar approach to differentiate the device layer (hardware and infrastructure devices), the communication layer (with physical and logical sub-layers), and the application or service layer (that runs applications and interacts with users). Some attack examples are then categorised according to the proposed taxonomy, with cases from consumer, commercial, and industrial IoT use cases.

Although multiple works have proposed taxonomies based on a set of attack cases, existing taxonomies are not built using the same global attack set \cite{Conti2021, Xenofontos2022}. Since only a subset of the complete set of attacks is used, published taxonomies fall short of proposing a global view of the attack-type universe. Some cyber-thread frameworks, like Cyber Kill Chain (CKC) \cite{LockheedMartin2011} and Mitre ATT\&CK \cite{Strom2020}, follow the perspective from the attacker’s view, instead of the network focused point of view. These frameworks then proceed to describe the multiple steps that each attack must perform to achieve the attack objective.

Mitre ATT\&CK includes a knowledge base of tactics, techniques and procedures that describe and document real attacks. The most common knowledge base, Enterprise ATT\&CK, was targeted at the enterprise environment and includes Windows, Linux and MacOS, Cloud and mobile devices. Another knowledge base, ATT\&CK ICS, was created specifically for Industrial Control Systems environments \cite{Alexander2020}.

In the knowledge base, the attacker’s goals i.e. “the reason for performing an action”, are categorised ranging from reconnaissance, initial access, execution, persistence, credential access, command and control, and exfiltration among others. For each attacker’s goal, a tactic, or a set of techniques are described as "how an adversary achieves a tactical objective". The ATT\&CK ICS knowledge base includes the following twelve tactics: initial access, execution, persistence, privilege escalation, evasion, discovery, lateral movement, collection, command and control, inhibit response function, impair process 

\noindent\scalebox{0.775}{
\begin{threeparttable}

\caption{Alphabetical list of industrial malicious datasets with references. Note that the third column refers to the year of publication of the dataset.}
\label{tab:32datasets}

\renewcommand{\arraystretch}{1.1}

\begin{tabular}{|c|p{5.7cm}|l|p{7cm}|p{1.3cm}|}\hline\rowcolor[HTML]{DEEAF6} 
\textbf{No.}  & \textbf{Dataset Name}   & \textbf{Year} & \textbf{Application   scenario}                     & \textbf{Ref.}    \\ \hline
1            & 2017QUT\_DNP3                            & 2021          & Critical   Infrastructure (Power)         &             \cite{Rodofile2018}~\tnote{a}                  \\
\rowcolor[HTML]{DEEAF6} 
2            & 2017QUT\_S7 (Myers)                      & 2017          & Industrial   Systems (Mining)             &          \cite{Myers2018}                     \\
3            & 2017QUT\_S7comm (Rodofile)               & 2017          & Industrial Systems (Mining)               &    \cite{Rodofile2017}~\tnote{b}                           \\\rowcolor[HTML]{DEEAF6} 
4            & A Industry 4.0   Production Line         & 2023          & Industrial   Systems (PLC)                &            \cite{LeiShi2023}                    \\
5            & BATADAL                                  & 2017          & Critical Infrastructure (Water)           &              \cite{Taormina2018}                 \\\rowcolor[HTML]{DEEAF6} 
6            & CIC Modbus 2023                          & 2017          & Industrial Systems (Modbus)               &                \cite{Boakye-Boateng2023}~\tnote{c}               \\
7            & Control Logic   Injection                & 2023          & Industrial   Systems (PLC)                &                     \cite{Yoo2019}          \\\rowcolor[HTML]{DEEAF6} 
8            & SANS Cyber City                          & 2013          & Critical   Infrastructure (Power)         &  online\tnote{d}  \\
9            & Cyber-Security   Modbus ICS              & 2019          & Industrial Systems (Modbus)               &     \cite{Frazao2019}                           \\\rowcolor[HTML]{DEEAF6} 
10           & DNP3 Intrusion   Detection               & 2016          & Industrial   Systems (DNP3)               &            \cite{RadoglouKelli2022}                    \\
11           & Edge-IIoT                                & 2022          & IoT and IIoT                              &                     \cite{Ferrag2022}           \\\rowcolor[HTML]{DEEAF6} 
12           & Electra Modbus   \& S7comm               & 2019          & Industrial   Systems (Railway, ICS \& S7) &                 \cite{Perales2019}               \\
13           & EPIC                                     & 2021          & Critical   Infrastructure (Power)         &      \cite{iTrust2021}                          \\\rowcolor[HTML]{DEEAF6} 
14           & HAI                                      & 2021          & Critical   Infrastructure (Power)         &                    \cite{Shin2021}            \\
15           & HDGM                                     & 2019          & Industrial   Systems (ICS)                &           \cite{Li2019}                     \\\rowcolor[HTML]{DEEAF6} 
16           & \parbox{4cm}{ICS Gas Pipeline   \& Water Storage Tank} & 2011          & Critical   Infrastructure (Gas \& Water) &                       \cite{Morris2011}         \\
17           & ICS Gas Pipeline                         & 2013          & Critical   Infrastructure (Gas)           &         \cite{Beaver2013}                       \\\rowcolor[HTML]{DEEAF6} 
18           & ICS New Gas   Pipeline                   & 2015          & Critical   Infrastructure (Gas)           &              \cite{Morris2015}                  \\
19           & ICS Power   System                       & 2014          & Critical   Infrastructure (Power)         &            \cite{Pan2015}                    \\\rowcolor[HTML]{DEEAF6} 
20           & IEC 60870-5-104                          & 2020          & Critical   Infrastructure (Smart Grid)    &            \cite{RadoglouRompolos2022}                    \\
21           & IEC 61850   Security                     & 2019          & Critical   Infrastructure (Smart Grid)    &               \cite{Biswas2019}                 \\\rowcolor[HTML]{DEEAF6} 
22           & ISOT                                     & 2022          & Military                                  &                \cite{Ahmed2023}                \\
23           & Lemay Modbus                             & 2016          & Industrial   Systems (Modbus)             &   online\tnote{e}         \\\rowcolor[HTML]{DEEAF6} 
24           & Maynard SCADA                            & 2018          & Industrial   Systems (SCADA)              &             \cite{Teixeira2018}             \\
25           & Modbus TCP SCADA   \#1                   & 2018          & Industrial   Systems (SCADA)              &                    \cite{Frazao2019}            \\\rowcolor[HTML]{DEEAF6} 
26           & Realtime ICS   SCADA                     & 2021          & Industrial   Systems (SCADA)              &                  \cite{Mubarak2021}              \\
27           & SWaT                                     & 2017          & Critical   Infrastructure (Water)         &    online\tnote{f}  \\\rowcolor[HTML]{DEEAF6} 
28           & TON\_IoT                                 & 2020          & IoT and IIoT                              &                   \cite{Alsaedi2020}             \\
29           & WADI                                     & 2017          & Critical   Infrastructure (Water)         &  online\tnote{g}    \\\rowcolor[HTML]{DEEAF6} 
30           & \parbox{4cm}{WUSTL-IIOT-2018 ICS (SCADA)}           & 2018          & Industrial   Systems (SCADA)              &  \cite{Teixeira2018}                              \\
31           & WUSTL-IIOT-2021                          & 2021          & Industrial Internet of Things (IIoT)                                      &                            \cite{Zolanvari2021}    \\\rowcolor[HTML]{DEEAF6} 
32           & X-IIOTID                                 & 2020          & Industrial Internet of Things (IIoT)                                       &                       \cite{Al-Hawawreh2021}        \\\hline
\end{tabular}

\begin{tablenotes}\small
 \item[a]{\url{https://github.com/qut-infosec/2017QUT_DNP3 }}
 \item[b]{\url{https://github.com/qut-infosec/2017QUT_S7comm}}
 \item[c]{\url{https://www.unb.ca/cic/datasets/modbus-2023.html }}
 \item[d]{\url{https://www.sans.org/mlp/holiday-challenge/2013/}}
 \item[e]{\url{https://github.com/antoine-lemay/Modbus_dataset}}
 \item[f]{\url{https://drive.google.com/drive/folders/1ABZKdclka3e2NXBSxS9z2YF59p7g2Y5I?usp=sharing}}
 \item[g]{\url{https://drive.google.com/drive/folders/11XWMQruwxFvlVEiqNgZ1mxVw-c-5xSmR?usp=sharing}}
   \end{tablenotes}
  
\end{threeparttable}
}

\clearpage

 \noindent control, and impact. Procedures describe the specific implementation of a technique and mitigation examples are presented to prevent an attack from being successful.

The CKC framework proposes seven sequential steps attackers perform to reach their goal:
\begin{itemize}
    \item Reconnaissance: Identify the Targets
   \item Weaponisation: Prepare the Operation
   \item Delivery: Launch the Operation
   \item Exploitation: Gain Access to Victim
   \item Installation: Establish Beachhead at the Victim
   \item Command \& Control: Remotely Control the Implants
   \item Actions on Objectives: Achieve the Mission’s Goal
\end{itemize}

Tables~\ref{tab:attacktableA} and~\ref{tab:attacktableB} present a summary of the 32 identified open access industrial datasets mapped to the CKC framework. We decided to map these attacks to the CKC framework due to its simple seven step chain, where each attack was mapped to one of the steps. Within the CKC steps, the Weaponisation step, when malware or malicious payload are created, and the Delivery step, when the malicious payload is delivered by social engineering, phishing, USB or web, do not contain attacks that are referenced in the 32 datasets documentation, as a result they are greyed out in the table. 
The dataset mapping presents 22 further subheadings of the CKC seven steps to help illustrate the technical focus or industrial protocol of the 97 attacks based on analysis of the dataset documentation. The subheadings in each CKC step are presented in attack chronological order unless the attacks are asynchronous in which case they are in alphabetical order.

\input{latexAttackTable}

\subsection{Complexity Analysis}
\label{sec:42}
The dataset complexity analysis was limited to such datasets that turned out to be ML-ready, i.e., properly labelled datasets which can be classified by ML algorithms without extensive pre-processing. Just 17 of the 32 sets provided at least one file of the network traffic data in the form of a labelled CSV or ARFF file. The other datasets solely consist of unlabelled traffic data, usually using PCAP format. Data in PCAP format cannot be regarded as ML-ready as it requires several pre-processing steps, some of which are not easily comparable or standardisable as they depend on the user-specific research intentions, especially the labelling of attacks. Table~\ref{tab:cs-results} summarises the statistics collected in the dataset analysis workflow described in Section~\ref{sec:35}. In addition to the name of the ML-ready data record, the table contains the name of the analysed file. This is either the one prepared and recommended for ML tasks by the creators of the dataset or, in several cases, a random selection of a labelled file. Column 3 of the table shows the dataset file format. The ML-ready files mainly include CSV files (14 out of 17), but also two in ARFF format and one Microsoft Excel (XLSX) file. The table then specifies the number of data points within the referenced file, whereas the number varies between one thousand and 16 million data points. The number of features is given in column five and ranges between five and 130. Interestingly, there are three datasets with fewer than ten features, which nevertheless show very different characteristics. 

Finally, the last two columns indicate the imbalance ratio ($IR$) and the average complexity score (Avg. $CS$). The $IR$ is commonly calculated by dividing the number of data points of the majority class, i.e., the class with the most data points, by the number of data points in the minority class. i.e., 
\begin{equation}
    IR =  \frac{|\textrm{majority class}|}{|\textrm{minority class}|}
\end{equation}
Note that across all datasets, the benign network traffic data points can always be considered the majority class and, conversely, the malicious traffic represents the minority class. Consequently, a value of $IR=1$ indicates a perfect balance of both classes, whereas large $IR$ values point towards high imbalance and only a few data points of malicious traffic, respectively. The 17 datasets exhibit $IR$ values in between the full range from 1 to 99 with an average $IR$ of 20.8. While 13 datasets have an $IR$ value below this average, only 4 datasets turn out to be heavily imbalanced with an $IR$ above 30. 
The calculation of the average complexity scores ($CS$) was performed by use of the Python Problexity module. The score is determined as the mean value of 22 distinct complexity measures and follows the explanations in Section~\ref{sec:35} above. The average $CS$ score resides in the interval $[0;1]$ and large values correspond to high complexity within the data. From Table 6, we understand that there is one dataset (DNP3 Intrusion Detection) with a $CS$ below 0.1. Complexity values between 0.1 and 0.2 are not discovered. Six datasets have a $CS$ in the range of 0.2 to 0.3, and seven lie between 0.3 and 0.4. The remaining three ML-ready datasets show an average complexity score in the range of 0.4 to 0.5. None of the considered datasets obtains a $CS$ value above 0.5. The average complexity  among the ML-ready datasets is around 0.323.

\begin{table}[htbp]

\caption{Table of statistics and avg. complexity for the identified ML-ready datasets.}
\scalebox{0.69}{
\begin{tabular}{p{3.8cm}p{6.2cm}p{1.5cm}p{1.55cm}p{1.3cm}p{1.2cm}p{1.1cm}}
\rowcolor[HTML]{DEEAF6} 
\textbf{Dataset   shortcut}             & \textbf{Selected   file or Comment}                                              & \textbf{File format} & \textbf{\# Data points} & \textbf{\# features} & \textbf{IR} & \textbf{Avg. CS} \\ \hline
2017QUT\_S7comm (Rodofile)              & AttackLog-Labelled.xlsx                                                          & xlsx                 & 3671                    & 7                    & 11.50       & 0.368            \\
\rowcolor[HTML]{DEEAF6} 
A Industry 4.0 Production Line          & Dataset A\_Real System Data\_Denial of Service attack.csv                        & csv                  & 5152                    & 61                   & 1.04        & 0.321            \\
DNP3 Intrusion Detection                & {20200516\_DNP3\_info\_UOWM \_DNP3\_Dataset\_Slave\_02{.}pcap 60DNP3\_FLOWLABELED.csv} & csv                  & 1112                    & 101                  & 1.00        & 0.075            \\
\rowcolor[HTML]{DEEAF6} 
Edge-IIoT                               & DNN-EdgeIIoT-dataset.csv                                                         & csv                  & 2219201                 & 62                   & 3.44        & 0.284            \\
Electra Modbus \& S7comm                & electra\_modbus.csv                                                              & csv                  & 16289277                & 10                   & 5.67        & 0.341            \\
\rowcolor[HTML]{DEEAF6} 
HAI                                     & hai21.03/test2.csv                                                               & csv                  & 118801                  & 83                   & 32.33       & 0.371            \\
HDGM                                    & train\_data.csv + test\_data.csv                                                 & csv                  & 3890                    & 78                   & 1.00        & 0.479            \\
\rowcolor[HTML]{DEEAF6} 
ICS Gas Pipeline                        & GasPipelineMulticlasCommand InjectionV2.csv                                       & csv                  & 28344                   & 26                   & 99.00       & 0.251            \\
ICS Gas Pipeline \& WaterStorage   Tank & gas\_final.arff                                                                  & arff                 & 97019                   & 25                   & 32.33       & 0.440            \\
\rowcolor[HTML]{DEEAF6} 
ICS New Gas Pipeline                    & IanArffDataset.arff                                                              & arff                 & 274628                  & 19                   & 3.57        & 0.430            \\
IEC 60870-5-104                         & 20200606\_UOWM\_IEC104\_Dataset \_c\_rp\_na\_1\_iecserver5.pcap\_Flow.csv         & csv                  & 33937                   & 83                   & 99.00       & 0.274            \\
\rowcolor[HTML]{DEEAF6} 
ISOT                                    & 1553\_logic\_injection\_attack6.csv                                              & csv                  & 1004                    & 51                   & 19.00       & 0.298            \\
TON\_IoT                                & IoT\_Fridge.csv                                                                  & csv                  & 587076                  & 5                    & 5.67        & 0.397            \\
\rowcolor[HTML]{DEEAF6} 
WADI                                    & WADI\_attackdataLABLE.csv                                                        & csv                  & 172803                  & 130                  & 15.67       & 0.364            \\
WUSTL-IIOT-2018\_ICS (SCADA)            & wustl-scada-2018.csv                                                             & csv                  & 7037983                 & 6                    & 15.67       & 0.211            \\
\rowcolor[HTML]{DEEAF6} 
WUSTL-IIOT-2021                         & wustl-ehms-2020\_with\_attacks\_categories.csv                                   & csv                  & 16317                   & 44                   & 6.69        & 0.211            \\
X-IIOTID                                & X-IIoTID\_dataset.csv                                                            & csv                  & 820833                  & 68                   & 1.04        & 0.367           
\end{tabular}
}
\label{tab:cs-results}
\end{table}

\subsection{Feature Analysis}
\label{sec:43}
The average complexity scores reported in Table~\ref{tab:cs-results} above show quite some variation in complexity between datasets. To better understand the dataset characteristics associated with this variation, we examine a relatively high complexity, a medium complexity, and a low complexity dataset in more detail. Figure 4 below shows feature importance graphs based on MI for the three datasets. 
\begin{figure}[htbp]
    (a)\includegraphics[width=0.945\textwidth]{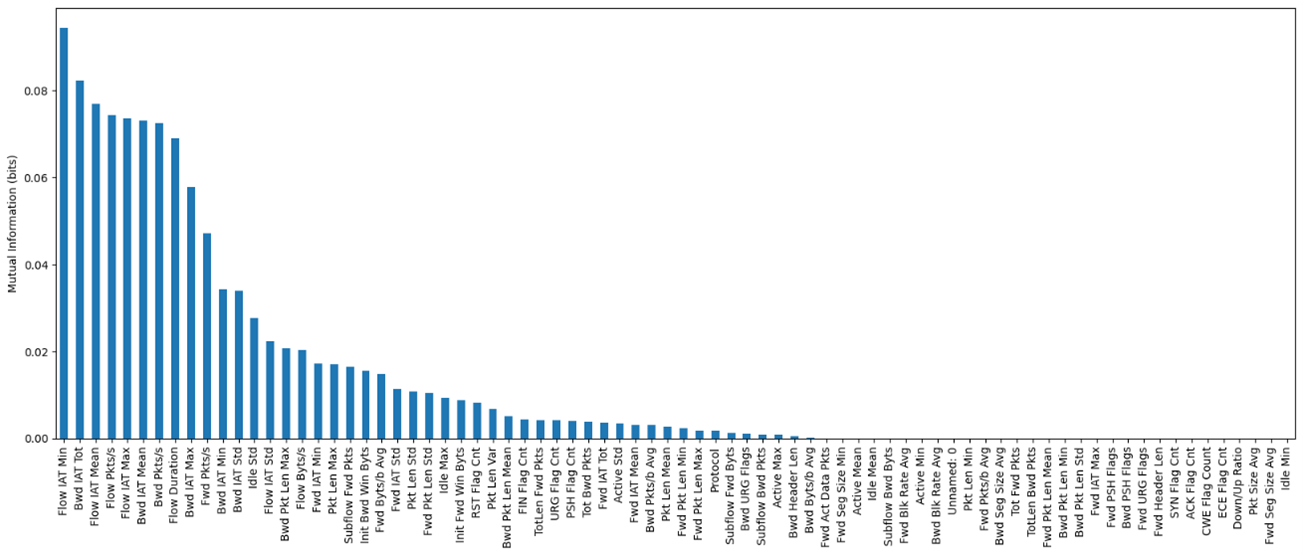}\\
    (b)\includegraphics[width=0.95\textwidth]{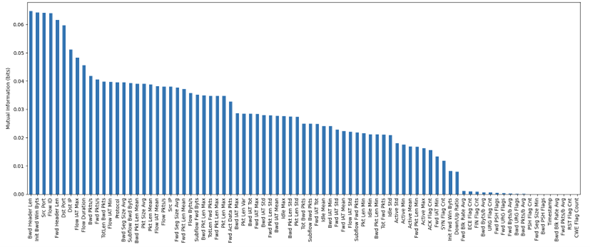}\\
    (c)\includegraphics[width=0.95\textwidth]{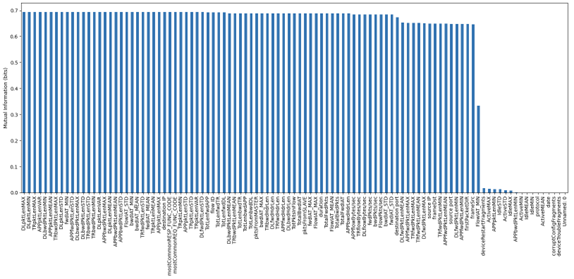}
    \caption{Feature importance plots for (a) high complexity dataset HDGM (Complexity Score 0.479), (b) medium complexity dataset IEC 60870-5-104 (Complexity Score 0.274), and (c) low complexity dataset DNP3 (Complexity Score 0.075). }
    \label{fig:FeatureAnalysis01}
\end{figure}

The HDGM dataset has the highest complexity score of 0.479. The feature importance plot for HDGM, Figure~\ref{fig:FeatureAnalysis01}(a), shows low MI values of less than 0.085 bits for the top five features, with many of the other features having MI values close to zero. Table~\ref{tab:mostimportant} shows feature plots for the five most important features of the HDGM dataset. All five features display large or almost total class overlap with no clear class boundaries evident on an individual dimension basis. 

The IEC 60870-5-104 dataset was selected as the medium complexity dataset with a complexity score of 0.274. Figure~\ref{fig:FeatureAnalysis01}(b), shows higher MI values for a large number of features. Table~\ref{tab:mostimportant} shows feature plots for the five most important features of the IEC 60870-5-104 dataset. Feature 1 has the highest MI value of 0.065 bits and a class overlap is visible between the benign and malicious classes.

The DNP3 dataset has the lowest complexity score of 0.075. Figure~\ref{fig:FeatureAnalysis01}(c), shows high MI values of close to 0.7 for many of the features. Table~\ref{tab:mostimportant} shows feature plots for the five most important features of the DNP3 dataset. While class overlap is evident in some features, the benign class often has quite a well-defined class boundary.

\begin{table}[htp]
\centering
\renewcommand{\arraystretch}{1.0}
\begin{tabular}{c|p{0.25\textwidth}p{0.25\textwidth}p{0.25\textwidth}}
  & \multicolumn{1}{c}{\textbf{HDGM}} &   \multicolumn{1}{c}{\textbf{IEC 60780-5-104}} &  \multicolumn{1}{c}{\textbf{DNP3}}\\
\textbf{No.} & \multicolumn{1}{c}{\textbf{$CS = 0.479$}} &   \multicolumn{1}{c}{\textbf{$CS= 0.274$}} &  \multicolumn{1}{c}{\textbf{$CS =0.075$}} 
\\\hline\hline

01 & 	 
\includegraphics[width=0.2\textwidth,height=0.175\textwidth]{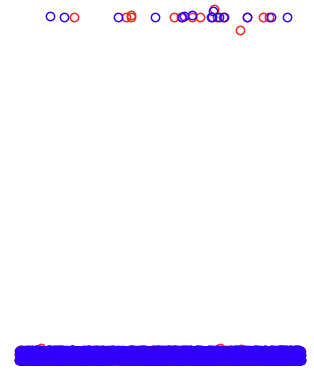} {\footnotesize Flow IAT Min} &  \includegraphics[width=0.2\textwidth,height=0.175\textwidth]{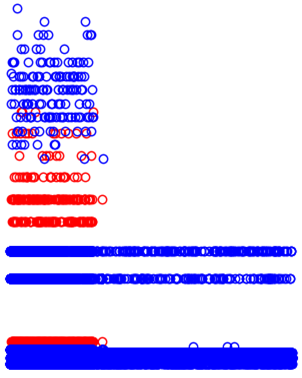} {\footnotesize Bwd Header Len}& 
\includegraphics[width=0.2\textwidth,height=0.175\textwidth]{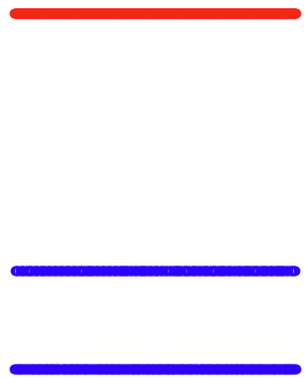} {\footnotesize DLpktLenMAX}\\\hline

02 & \includegraphics[width=0.2\textwidth,height=0.175\textwidth]{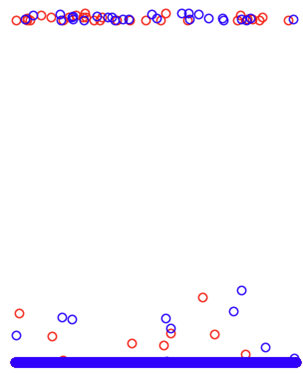} {\footnotesize Bwd IAT Tot }
&  \includegraphics[width=0.2\textwidth,height=0.175\textwidth]{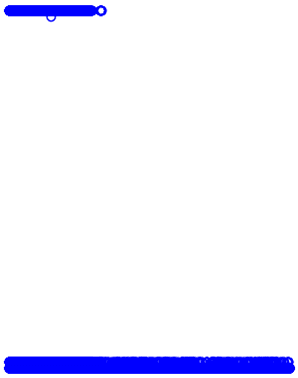} \quad {\footnotesize Init Bwd Win Byts }	 
&\includegraphics[width=0.2\textwidth,height=0.175\textwidth]{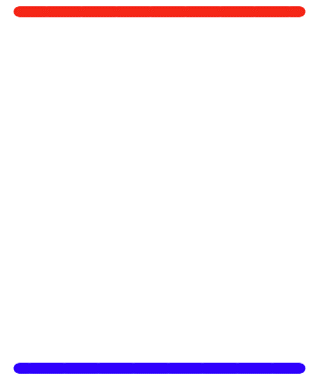} \quad{\footnotesize DLpktLenMIN }\\\hline

03 & \includegraphics[width=0.2\textwidth,height=0.175\textwidth]{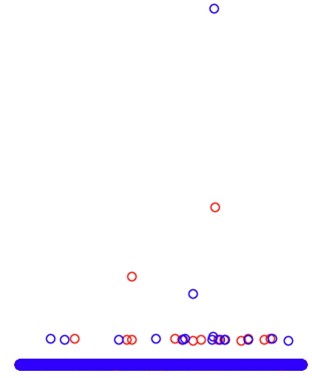}{\footnotesize Flow IAT Mean }	 
 &  \includegraphics[width=0.2\textwidth,height=0.175\textwidth]{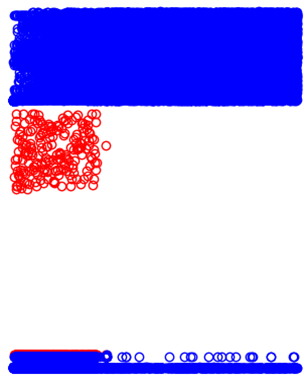} \quad {\footnotesize Src Port }	 
&\includegraphics[width=0.2\textwidth,height=0.175\textwidth]{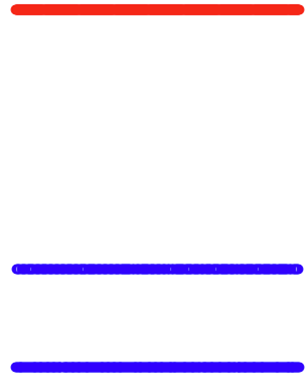} {\footnotesize TRpktLenMAX }\\\hline

04& \includegraphics[width=0.2\textwidth,height=0.175\textwidth]{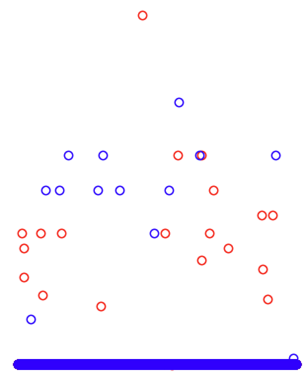}{\footnotesize Flow Pkts/s  }	 
 &  \includegraphics[width=0.2\textwidth,height=0.175\textwidth]{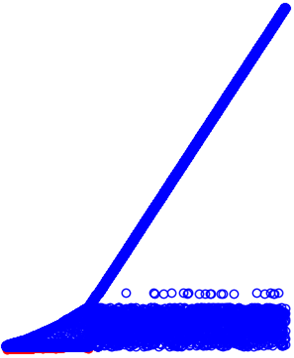} {\footnotesize Flow ID	  }
&\includegraphics[width=0.2\textwidth,height=0.175\textwidth]{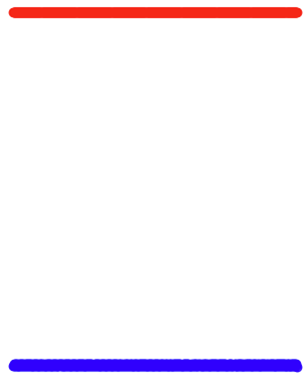} {\footnotesize APPpktLenVAR } \\\hline

05 & \includegraphics[width=0.2\textwidth,height=0.175\textwidth]{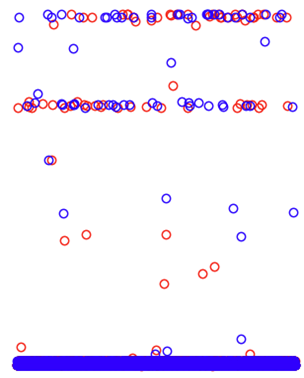} {\footnotesize Flow IAT Max }	 
&  \includegraphics[width=0.2\textwidth,height=0.175\textwidth]{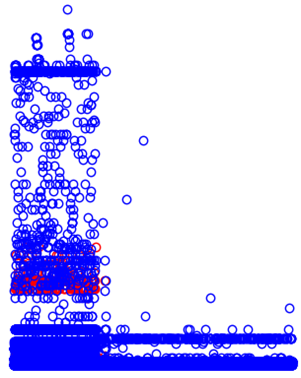} {\footnotesize Fwd Header Len }	 
&\includegraphics[width=0.2\textwidth,height=0.175\textwidth]{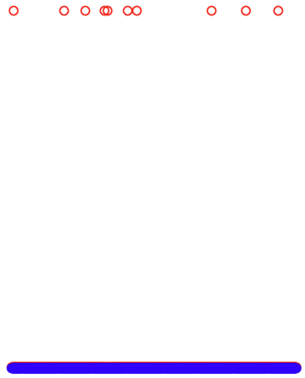} {\footnotesize DLbwdPktLenMIN }\\\hline

\end{tabular}
\caption{Individual feature plots showing the distribution of benign (blue) and malicious (red) traffic for the five most important features of each dataset. The horizontal axis of each plot is the row (or data point) index in the analysed CSV file and the vertical axis is the feature value on a linear scale from the feature’s minimum to its maximum value.}
\label{tab:mostimportant}
\end{table}

% \newpage

% \input{latexFeatureTable}

% \newpage

\section{Discussion}
\label{sec:discussion}

In this section, we discuss the results obtained for the attack categorisation, complexity analysis and feature analysis.

\subsection{Attack Categorisation}

The source dataset documentation attack descriptions were reviewed and mapped to the Lockheed Martin CKC attack framework. CKC provides seven summary categories and we then developed 22 subheadings to provide greater granularity. The subheadings were designed to reflect the IIoT/OT environment and protocols in the datasets.

The attack distribution across the multiple steps is not uniform. While most of the datasets contain Exploitation instances, the other types of attacks namely Reconnaissance, Installation, Command \& control, and Actions on objectives, are not present in all datasets.

The placement of attacks in the subheadings is a compromise and open for debate as attacks may be legitimately placed in several CKC steps in the attack framework, e.g. DDoS might be placed in Exploitation or Command \& Control, but the main intent is to inform researchers as to which datasets may be most relevant for their research.

\subsection{Complexity Analysis}
Following the workflow presented in Section~\ref{sec:35}, we examined 17 of the 32 selected datasets regarding their classification complexity. The results are summarized in Table~\ref{tab:cs-results} and described in Section~\ref{sec:42}. The analysis uses the Python module Problexity, which condenses 22 separate complexity measures to a single average complexity score (CS). For each data set, we provide the CS and the associated IR, as well as information on the number of features and data points, to give the readers valuable first information to judge the suitability of a dataset for their purposes. It must be mentioned that our results only provide an initial CS estimate and should not be misunderstood as the actual CS of the data set. This is due to limiting the calculation of the CS to a representative sample of 1000 data points per dataset while, at the same time, maintaining the original dataset IR in the sample. Likewise, the fact that IR and CS values were only determined based on a binary class understanding of benign and malicious traffic must be considered. The reason for these simplifications is to reduce the computational effort otherwise required to a reasonable level. In this respect, our information serves for informed dataset selection on the way to targeted ML experimentation.

When looking at the IR values of the analysed datasets, we observe an average value of 20.8 over the 17 data sets and only 4 datasets of relatively large class imbalance. Conversely, there are 9 datasets with a benign to malicious traffic ratio below 10:1. This raises the question of whether such datasets are designed realistically because certainly in a typical OT environment a very small proportion of network traffic log is malicious.  This could indicate an unintended design bias caused by a focus on the development of ML based intrusion detection algorithms. Since highly unbalanced data sets generally pose difficulties for ML and the trained models are likely to develop one-sided classification capabilities, data engineers might want to rule this issue out from the very beginning and pay attention to a certain balance in the composition of the data sets. However, the low IRs can of course also have other reasons, such as focussing on a time window with a particularly high number of attacks or looking at a particularly vulnerable system. In any case, it is useful for the users of the data sets to think about the realism of the data to be able to assess the capabilities of the ML algorithms trained with it when applied to new unknown data and real-world application domains.

The dataset of the lowest classification complexity appears to be DNP3 Intrusion Detection with a $CS$ value of 0.075. While the dataset contains about 9 types of attacks (see Tables~\ref{tab:attacktableA} and\ref{tab:attacktableB}), the Imbalance Ratio $IR = 1$ indicates a perfect balance between benign and malicious traffic. Yet such a relationship between low $IR$ and low $CS$ value is by no means the rule. Taking into account the HDGM dataset, one observes the contrary. HDGM comes with an $IR$ of 1 but receives the highest value $CS=0.479$ value among the considered datasets. Also the four most imbalanced datasets ($IR >30$) do have quite different CS values and only ICS Gas Pipeline \& Water Storage Tank belongs to the three most complex datasets. In summary, the values of $IR$ and $CS$ appear to not exhibit any correlation. The reason for this is twofold. Firstly, the complexity of a dataset cannot be described by class imbalance (i.e., the $IR$) alone, as many features influence the degree of difficulty for ML purposes. In this respect, our observations are in line with the fact that dataset complexity depends on more aspects than only class imbalance \cite{Santos2023}. Secondly, as pointed out earlier the average $CS$ represents the average of 22 metrics and disguises therefore the reasons for the measured dataset complexity to some degree. This complexity may or may not lie in the imbalance and thus no mandatory correlation should be expected here. A more detailed analysis is outside the scope of this paper. The investigation of individual complexity characteristics and their correlations will be analysed in more detail in future work.

    \subsection{Feature Analysis}
The feature analysis is performed on the three datasets: DNP3, IEC 60870-5-104 and HDGM; according to their complexity score. After analysing the feature importance, the five most important features of each dataset are considered in more detail.

\subsubsection{DNP3}
Figure~\ref{fig:FeatureAnalysis01}(c) is the feature importance for the DNP3 data. The DNP3 data was captured from the testbed as PCAP files and converted into CSV files using a custom DNP3 Python parser. The CSV file contains 101 features of which four are categorical and 97 are numerical.  The numerical features are mostly DNP3 flow statistics but also include other numerical features such as source and destination port numbers. A clear characteristic of Figure~\ref{fig:FeatureAnalysis01}(c) is the high mutual information for most features. There are 84 features with over 0.6 bits of mutual information. We made no effort to determine the orthogonality of each of these features however Figure~\ref{fig:FeatureAnalysis01}(c)suggests that many features may be correlated, and that the dataset may benefit from the use of a dimensionality reduction technique such as Principal Component Analysis (PCA). For a binary classification problem with $IR = 1.00$, it is likely that just a small number of features would perform as well as the use of all 101 features although we have not tested this.
Table~\ref{tab:mostimportant} shows a more detailed analysis of the five most important features. These features show clear class separation with no visible class overlap. All these features are related to IP packet length and indicate that the malicious IP packets had longer lengths than the benign packets. 
% Features 9 and 10 show clear class overlap however their high mutual information indicates that the classes are separable.

\subsubsection{IEC 60870-5-104}
Figure~\ref{fig:FeatureAnalysis01}(b) is the feature importance for the IEC 60870-5-104 data. The CSV file contains 83 features of which 4 are categorical and 79 are numerical. Figure~\ref{fig:FeatureAnalysis01}(b) shows lower mutual information for each feature when compared with the DNP3 data. The 5 features have approximately 0.06 bits of mutual information each. Table~\ref{tab:mostimportant} shows a more detailed analysis of the five most important features. All the features show a clear dependence on the X-axis value. This is an artefact of the data collection process where benign and malicious traffic are grouped in the data files. ML researchers should make sure to avoid approaches where a classifier includes time or sample number when making predictions. Feature 4 shows a diagonal line indicating a correlation between IP flows and time however this a just an artifact of the categorical label encoding process that we used.  For example, for Destination IP addresses, the label encoder simply takes each new IP address encountered in the dataset and assigns it the next numerical value.

\subsubsection{HDGM}
Figure~\ref{fig:FeatureAnalysis01}(a) is the feature importance for the HDGM data. The CSV file contains 78 features all of which are numerical.  This is the dataset with the highest complexity and as expected has the lowest mutual information across all features. Almost half of the features have mutual information close to zero bits indicating that they could be dropped from an ML model without impacting performance. The selection of features for an ML model has great impact on the classification performance, since some typical features like real IP addresses or even artificial features like sample numbering, might cause an artificial improvement in the classification score \cite{Fernandes2023}. Table~\ref{tab:mostimportant} shows a more detailed analysis of the five most important features. All five features appear to show total class overlap with no discernible class boundaries. Nonetheless, a complexity value of 0.479 indicates that standard ML algorithms should be able to classify with reasonable performance and we could expect a density of manifolds view at higher dimensions to show evident class boundaries.

\section{Summary and Conclusion}
\label{sec:conclusion}

\subsection{Conclusion}
The integration of the IIoT with OT has ushered in a new era of standardised data accessibility and enhanced process transparency. While beneficial in industrial use cases, it has also introduced novel cybersecurity challenges that need to be addressed. In response, AI and ML are increasingly being leveraged to monitor networks for potential threats, marking a significant area of ongoing research to enhance existing IDS or similar technologies. In doing so, AI specialists, researchers and practitioners are depending on publicly available datasets to test, train, and refine their models, highlighting the growing demand for such datasets in research and practice.

This research examines publicly available datasets used in ML research published between 2019 and 2023. It includes an assessment of the types of attacks featured, an analysis of metadata, and evaluations of both statistical and complexity characteristics. Whenever possible, each data set is described in detail to provide researchers with (meta-)data, aiding them in selecting the most suitable dataset for their specific (research) needs.

There is an abundance of datasets available for research using ML algorithms for the detection of malicious network traffic. We identified 32 datasets relevant to OT networks and studied them in detail. Of these, the Exploitation attack type is the most commonly present on the datasets. Other types of attacks, from the early Reconnaissance up to the late Command and Control and Action on objectives, are not present on all datasets, which leads to the question of their relevance for an intrusion attack on an IIoT/OT system. These last step attacks can be damaging to a system and hence they should deserve more attention so that ML algorithms are trained to detect them.

We then proceed to study the complexity of selected datasets, limited to the analysis of datasets which can be considered ready for ML applications from the get-go. Out of 32 datasets, 17 are deemed ML-ready, meaning they are properly labelled and can be used in ML algorithms without extensive pre-processing. These datasets consist normally of network traffic data in labelled CSV or ARFF files. The other datasets, mainly in PCAP format, are not considered ML-ready as they require significant pre-processing and lack standardised labelling. For the 17 datasets, we show the number of features and datapoints, the imbalance ratio, and the average complexity score. There is no apparent correlation between IR and CS values. This lack of correlation can be attributed to the fact that dataset complexity cannot be solely determined by class balance; it is influenced by various factors. Additionally, the average CS, based on 22 metrics, usually obscures the specific reasons for a dataset's complexity. 

We finally analysed the Feature Analysis for three exemplary datasets with different complexity score levels. Since many datasets contain multiple data files, the features present in each file for the same dataset vary significantly. We observed that some files contained features that were artifacts of the experiment (ex: the IP address or sequential number scheme) that could lead to artificially good results when used to train and test a ML algorithm. Finally, we found a wide variation in the mutual information of features indicating that AI performance testing would be very dependent on the dataset chosen.

\subsection{Recommendations \& Future Work}
The future of cybersecurity in IIoT/OT environments is undoubtedly supported by uninterrupted algorithmic and operational innovation, publicly available test data and even testbeds, as well as a deepened understanding of the intricate dynamics between technology, data, and security as well as the continuous question how to automate the detection of threats and threat actors. In the analysis of 32 public datasets, we realise that the foundation for research and practice has been established; however, it is still a long way to achieving full automation of (unknown) threat detection, even if AI/ML offers a glimpse into a potential future.
To drive forward the uptake, usefulness, and usability of datasets we derive the following five recommendations:
\begin{enumerate}
    \itemsep10pt
    \item \textbf{Address Challenges Faced by Dataset Users and Publishers:}

        A common challenge lies in the handling of datasets with numerous files, non-ML-ready formats, and missing labelling of benign and malicious traffic. Dataset publishers should prioritise formatting data in ML-friendly and commonly used formats like CSV or ARFF, as well as ensuring comprehensive, accurate labelling, enabling usage of the datasets from the get-go. By establishing guidelines for data format, including the pre-processing of more detailed and complex formats like PCAP into more accessible formats, ML results become comparable while maintaining data integrity. Whenever ML-friendly formats are not available, a standardised pre-processing pipeline should be kept in mind (see next recommendation).

        \item  \textbf{Standardise the Pre-Processing Pipeline for ML Practitioners:}

        We recommend the creation of a standardised pre-processing pipeline to aid in converting various data formats, especially PCAP files, into ML-ready forms. This pipeline should provide detailed steps and guidelines for pre-processing steps, data structures, and potential selection as well as configuration of pre-processing tools for easier ML application.

        \item  \textbf{Mitigate Risks of Uninformed Dataset Utilisation:}

        We advise against the use of certain data features that might lead to biased or inaccurate model training, such as source and destination IPs in many data sets. Often, these features are a result of individual captures or testbed specifics.  One solution is for the dataset publishers to mark features which can be excluded for ML approaches, and another is to recommend the removal to prevent artificially good performance.

        \item  \textbf{Choose Datasets Relevant to the Application Scenario:}

        The importance of selecting datasets that reflect the specific contextual conditions and attack types pertinent to the user's application must be stressed. Contextualising the application scenario ensures that the ML algorithms are trained on relevant and realistic scenarios, which might differ significantly in available data and features, used technologies and protocols, as well as attack types. Depending on the environment selected, IIOT or OT, the reader could choose the most suitable dataset according to the protocols and attack types present on the dataset itself, along with its complexity.

        \item  \textbf{Be Aware of Diverse Complexity When Testing and Comparing MA Algorithms:}

        For testing and comparing MA algorithms, it is useful to use a diverse range of datasets with varying levels of complexity. Selection of a wider range of features within one dataset can further make testing more comparable. In addition to better comparability, this approach might also lead to a more comprehensive assessment of the algorithm's capabilities and limitations in different scenarios.
\end{enumerate}
 
For future work and research, we plan to address some of the questions and open points derived from the above recommendations. Firstly, we want to focus on a detailed analysis of attack types contained in datasets, extending existing attack categorisations specifically for ML use. Secondly, future work should involve classification of attack types in existing datasets and analysing their evolution over time. This can help provide insights into the most prevalent threats and emerging trends in cybersecurity attacks. Also, there is a need to identify attack types that are currently underrepresented in openly available datasets, especially when aiding the development of more testbed/testing datasets that reflect the full spectrum of potential cybersecurity threats. Thirdly, we feel that there is an urgent need to develop a standardised pre-processing pipeline. By standardising data processing, researchers and practitioners can more easily access and use datasets, accelerating the development of effective intrusion detection solutions and allowing for easier comparison of ML algorithms and automated detection solutions. Lastly, we want to extend our complexity calculation methodology to refine dataset assessment. These efforts will focus on extending the complexity calculation methodology, by refining metrics such as Imbalance Ratio for larger datasets with a multitude of diverse features and introducing new state-of-the-art measures like, e.g., Hostility to provide a more nuanced understanding of dataset complexity \cite{Lancho2023}.

Linked to this effort, our analysis of the available OT datasets showed a significant variation in data complexity scores.  This variation has led us to consider two further research questions:
\begin{enumerate}
    \item What is the correlation, if any, between reported machine learning algorithm performance and dataset complexity score?
    \item How would the complexity scores change if features that appear to be artefacts of the testbeds are removed?
\end{enumerate}
 
We also observed that many of the datasets contained many sub-files. Dataset complexity and features often vary within these files therefore more work is required to characterise each dataset completely. Out random file selection approach is simply a view of a small portion of the dataset and therefore may not be representative of the entire dataset.

\section*{Acknowledgements}
From 04/2023 to 03/2027, the IoT Sustainability Lab will be funded by the Interreg VI "Alpenrhein-Bodensee-Hochrhein" (ABH) programme, whose resources are provided by the European Regional Development Fund (ERDF) and the Swiss Confederation. The funders had no role in study design, data collection and analysis, decisions to publish, or preparation of the manuscript. 

Further, the financial support from the Austrian Federal Ministry of Labour and Economy, the National Foundation for Research, Technology and Development and the Christian Doppler Research Association are gratefully acknowledged. 

The authors would also like to thank the Regional University Network – European University for connecting us. 

%% If you have bibdatabase file and want bibtex to generate the
%% bibitems, please use
%%
 \bibliographystyle{elsarticle-num-names} 
 %NL: \bibliography{cas-refs}
\bibliography{bibmaliciousdatasets}
%% else use the following coding to input the bibitems directly in the
%% TeX file.

% \begin{thebibliography}{00}

% %% \bibitem{label}
% %% Text of bibliographic item

% \bibitem{}

% \end{thebibliography}
\end{document}

%% file: latexAttackTable.tex
\begin{landscape}

\begin{table}[htbp]\centering
\caption{Part A -- Mapping of Industrial (OT) Malicious Traffic Datasets to the Cyber Kill Chain Framework.
}
\label{tab:attacktableA}
\scalebox{0.65}{\renewcommand{\arraystretch}{1.1}
\begin{tabular}{|cp{4cm}c|c|p{0.5cm}|p{0.5cm}|p{0.5cm}|p{0.5cm}|p{0.5cm}p{0.5cm}|l|l|l|l|l|l|l|l|l|l|p{0.5cm}|p{0.5cm}|c|c|c|c|p{0.5cm}|p{0.5cm}|p{0.5cm}|} \hline
& \multicolumn{3}{l}{ } & \multicolumn{1}{c}{ }  & \multicolumn{24}{c|}{ \textbf{Lockheed   Martin Cyber Kill Chain -- Seven steps}}  \\ \hline
                  &    &                      &                      & \multicolumn{4}{l|}{\cellcolor[HTML]{BDD6EE} \textbf{Reconnaissance}}                                                     & \multicolumn{2}{|c|}{ \cellcolor[HTML]{FBE4D5} }   & \multicolumn{10}{c|}{\cellcolor[HTML]{BDD6EE} \textbf{Exploitation}} & \multicolumn{2}{c|}{\cellcolor[HTML]{BDD6EE}  \textbf{Instal-}} & \multicolumn{4}{|l|}{\cellcolor[HTML]{BDD6EE}\textbf{Command }} & \multicolumn{3}{l|}{\cellcolor[HTML]{BDD6EE}\textbf{Actions on}} \\
&  &                      &                      & \multicolumn{4}{l|}{\cellcolor[HTML]{BDD6EE} \textbf{ }}& \multicolumn{2}{|c|}{ \cellcolor[HTML]{FBE4D5} }& \multicolumn{6}{l}{\cellcolor[HTML]{BDD6EE} \textbf{ }}                                                     & \multicolumn{4}{c|}{\cellcolor[HTML]{DEEAF6}\textbf{OT Protocols} }   & \multicolumn{2}{l|}{\cellcolor[HTML]{BDD6EE} \textbf{lation}}& \multicolumn{4}{|l|}{\cellcolor[HTML]{BDD6EE} \textbf{\& Control }}& \multicolumn{3}{|l|}{\cellcolor[HTML]{BDD6EE}\textbf{Objectives }}   \\
  \multirow{15}{*}{\textbf{No.}}& 
 \multirow{15}{*}{\textbf{Name of dataset}}&\multirow{15}{*}{\rotatebox[origin=c]{0}{\textbf{Year}}} &\multirow{8}{*}{\rotatebox[origin=c]{90}{\textbf{\# of attack types} }} & \multirow{4}{*}{\rotatebox[origin=c]{90}{TCP/UDP Port Scan }} & \multirow{4}{*}{\rotatebox[origin=c]{90}{Modbus / Scada Scan }} & \multirow{4}{*}{\rotatebox[origin=c]{90}{OS Fingerprint }} &\multirow{4}{*}{\rotatebox[origin=c]{90}{Vulnerability Scan }}  & \cellcolor[HTML]{FBE4D5}  &  \cellcolor[HTML]{FBE4D5}& \multirow{4}{*}{\rotatebox[origin=c]{90}{DoS /DDoS }}&\multirow{4}{*}{\rotatebox[origin=c]{90}{Injection-Protocol/Data}} &\multirow{4}{*}{\rotatebox[origin=c]{90}{Injection - SQL / XSS }} & \multirow{3}{*}{\rotatebox[origin=c]{90}{MITM }}&\multirow{4}{*}{\rotatebox[origin=c]{90}{PLC web service }} &\multirow{3}{*}{\rotatebox[origin=c]{90}{Replay}} &\multirow{3}{*}{\rotatebox[origin=c]{90}{DNP3 }} &\multirow{3}{*}{\rotatebox[origin=c]{90}{GOOSE }}&\multirow{3}{*}{\rotatebox[origin=c]{90}{S7 Comm }}&\multirow{4}{*}{\rotatebox[origin=c]{90}{SCADA/Modbus }} & \multirow{3}{*}{\rotatebox[origin=c]{90}{Backdoor }}&\multirow{3}{*}{\rotatebox[origin=c]{90}{Malware }}&\multirow{4}{*}{\rotatebox[origin=c]{90}{Brute Force }}&\multirow{4}{*}{\rotatebox[origin=c]{90}{Dictionary }} &\multirow{4}{*}{\rotatebox[origin=c]{90}{Malicious Insider }}&\multirow{3}{*}{\rotatebox[origin=c]{90}{Upload }}&\multirow{4}{*}{\rotatebox[origin=c]{90}{Exfiltration }}&\multirow{4}{*}{\rotatebox[origin=c]{90}{Tampering }}&\multirow{4}{*}{\rotatebox[origin=c]{90}{Ransomware }}\\
&& & & & &  & & \cellcolor[HTML]{FBE4D5}&\cellcolor[HTML]{FBE4D5} &  & & & & & & & & &  & & & &  & & & & & \\
 && & & & &  & & \cellcolor[HTML]{FBE4D5}&\cellcolor[HTML]{FBE4D5} &  & & & & & & & & &  & & & &  & & & & &\\
 & & & & & &  & & \cellcolor[HTML]{FBE4D5}&\cellcolor[HTML]{FBE4D5} &  & & & & & & & & &  & & & &  & & & & & \\
 && & & & &  & & \cellcolor[HTML]{FBE4D5}&\cellcolor[HTML]{FBE4D5}&  & & & & & & & & &  & & & &  & & & & &\\
 && & & & & & & \cellcolor[HTML]{FBE4D5}&\cellcolor[HTML]{FBE4D5}  &  & & & & & & & & &  & & & &  & & & & &\\
 && & & & & & & \cellcolor[HTML]{FBE4D5}&\cellcolor[HTML]{FBE4D5}  &  & & & & & & & & &  & & & &  & & & & &\\
 && & & & & & & \cellcolor[HTML]{FBE4D5}\multirow{-10}{*}{\rotatebox[origin=c]{90}{\textbf{Weaponisation} }}&\cellcolor[HTML]{FBE4D5}\multirow{-12}{*}{\rotatebox[origin=c]{90}{\textbf{Delivery} }} &  & & & & & & & & &  & & & &  & & & & &\\\hline
 1&\textbf{2017QUT\_DNP3} &  2021 &  6 &  •&   &   &   & \cellcolor[HTML]{FBE4D5}  &  \cellcolor[HTML]{FBE4D5} &  & •&  & •&  & •& •& •&  &  &  &  &  &  &  &  &  &  & \\
\rowcolor[HTML]{DEEAF6} 2&
\textbf{2017QUT\_S7 (Myer)}                     & 2017 & 21 &   &   &   &   & \cellcolor[HTML]{FBE4D5} & \cellcolor[HTML]{FBE4D5} &           & • &           & •         &           &           &           &           & •         & &   &   &   &   &   &   &   &   &                      \\
3&\textbf{2017QUT\_S7comm (Rodofile)}             & 2017 & 64 &   &   &   &   & \cellcolor[HTML]{FBE4D5} & \cellcolor[HTML]{FBE4D5} &           &  • &           & •         &           &           &           &           & •         &                                              &   &   &   &   &   &   &   &   &                      \\
\rowcolor[HTML]{DEEAF6} 4&
\textbf{A Industry 4.0 Production Line}         & 2023 & 1  &   &   &   &   & \cellcolor[HTML]{FBE4D5} & \cellcolor[HTML]{FBE4D5} & •         &   &           &           &           &           &           &           &           & •                                            &   &   &   &   &   &   &   &   &                      \\
5&\textbf{BATADAL}                                & 2017 & 14 &   &   &   &   & \cellcolor[HTML]{FBE4D5} & \cellcolor[HTML]{FBE4D5} &           &   &           & •         &           & •         &           &           &           & •                                            &   &   &   &   &   &   &   &   &                      \\
\rowcolor[HTML]{DEEAF6} 6&
\textbf{CIC Modbus 2023}                        & 2023 & 9  &   & • &   &   & \cellcolor[HTML]{FBE4D5} & \cellcolor[HTML]{FBE4D5} &           & • &           &           &           & •         &           &           &           & •                                            &   &   & • &   &   & • &   &   &                      \\
7&\textbf{Control Logic Injection}                & 2020 & 2  &   &   &   &   & \cellcolor[HTML]{FBE4D5} & \cellcolor[HTML]{FBE4D5} &           & • &           &           &           &           &           &           &           & •                                            &   &   &   &   &   &   &   &   &                      \\
\rowcolor[HTML]{DEEAF6} 8& 
\textbf{SANS Cyber City}                        & 2013 & 5  & • &   & • &   & \cellcolor[HTML]{FBE4D5} & \cellcolor[HTML]{FBE4D5} & •         & • &           & •         &           &           &           &           &           & •                                            &   &   &   &   &   &   &   &   &                      \\
9&\textbf{Cyber-Security Modbus ICS}              & 2019 & 4  &   &   &   &   & \cellcolor[HTML]{FBE4D5} & \cellcolor[HTML]{FBE4D5} & •         &   &           & •         &           &           &           &           &           & •                                            &   &   &   &   &   &   &   &   &                      \\
\rowcolor[HTML]{DEEAF6} 10&
\textbf{DNP3 Intrusion Detection}               & 2016 & 9  &   &   &   &   & \cellcolor[HTML]{FBE4D5} & \cellcolor[HTML]{FBE4D5} & •         & • &           & •         &           & •         & •         &           &           &                                              &   &   &   &   &   &   &   &   &                      \\
11&\textbf{Edge-IIoT}                              & 2022 & 14 & • &   & • & • & \cellcolor[HTML]{FBE4D5} & \cellcolor[HTML]{FBE4D5} & •         & • & •         & •         &           & •         &           &           &           & •                                            & • & • & • &   &   & • &   &   & •                    \\
\rowcolor[HTML]{DEEAF6} 12&
\textbf{Electra Modbus \& S7comm}               & 2019 & 7  &   & • &   &   & \cellcolor[HTML]{FBE4D5} & \cellcolor[HTML]{FBE4D5} & •         & • &           & •         &           & •         &           &           & •         & •                                            &   &   &   &   &   &   &   &   &                      \\
13&\textbf{EPIC}                                   & 2021 & 6  &   &   &   &   & \cellcolor[HTML]{FBE4D5} & \cellcolor[HTML]{FBE4D5} & •         & • &           &           &           &           &           & •         &           & •                                            &   &   &   &   &   &   &   &   &                      \\
    \rowcolor[HTML]{DEEAF6}     14&
\textbf{HAI}                                    & 2021 & 47 &   &   &   &   & \cellcolor[HTML]{FBE4D5} & \cellcolor[HTML]{FBE4D5} & •         & • &           & •         &           & •         &           &           &           & •                                            &   &   &   &   &   &   &   &   &                      \\
15&\textbf{HDGM}                                   & 2019 & 2  &   &   &   &   & \cellcolor[HTML]{FBE4D5} & \cellcolor[HTML]{FBE4D5} & •         &   &           & •         &           &           &           &           &           & •                                            &   &   &   &   &   &   &   &   &                      \\
\rowcolor[HTML]{DEEAF6}16& 
\textbf{ICS Gas Pipeline \& Water Storage Tank} & 2011 & 5  &   & • &   &   & \cellcolor[HTML]{FBE4D5} & \cellcolor[HTML]{FBE4D5} & •         & • &           &           &           &           &           &           &           & •                                            &   &   &   &   &   &   &   &   &                      \\\hline
\end{tabular}
}
\end{table}
\end{landscape}

\begin{landscape}

\begin{table}[htbp]\centering
\caption{Part B -- Mapping of Industrial (OT) Malicious Traffic Datasets to the Cyber Kill Chain Framework.}
\label{tab:attacktableB}
\scalebox{0.65}{\renewcommand{\arraystretch}{1.1}
\begin{tabular}{|cp{4cm}c|c||p{0.5cm}|p{0.5cm}|p{0.5cm}|p{0.5cm}|p{0.5cm}p{0.5cm}|l|l|l|l|l|l|l|l|l|l|p{0.5cm}|p{0.5cm}|c|c|c|c|p{0.5cm}|p{0.5cm}|p{0.5cm}|} \hline
& \multicolumn{3}{l}{ } & \multicolumn{1}{c}{ }  & \multicolumn{24}{c|}{ \textbf{Lockheed   Martin Cyber Kill Chain -- Seven steps}}  \\ \hline
                    &  &                      &                      & \multicolumn{4}{l|}{\cellcolor[HTML]{BDD6EE} \textbf{Reconnaissance}}                                                     & \multicolumn{2}{|c|}{ \cellcolor[HTML]{FBE4D5} }   & \multicolumn{10}{c|}{\cellcolor[HTML]{BDD6EE} \textbf{Exploitation}} & \multicolumn{2}{c|}{\cellcolor[HTML]{BDD6EE}  \textbf{Instal-}} & \multicolumn{4}{|l|}{\cellcolor[HTML]{BDD6EE}\textbf{Command }} & \multicolumn{3}{l|}{\cellcolor[HTML]{BDD6EE}\textbf{Actions on}} \\
&  &                      &                      & \multicolumn{4}{l|}{\cellcolor[HTML]{BDD6EE} \textbf{ }}& \multicolumn{2}{|c|}{ \cellcolor[HTML]{FBE4D5} }& \multicolumn{6}{l}{\cellcolor[HTML]{BDD6EE} \textbf{ }}                                                     & \multicolumn{4}{c|}{\cellcolor[HTML]{DEEAF6}\textbf{OT Protocols} }   & \multicolumn{2}{l|}{\cellcolor[HTML]{BDD6EE} \textbf{lation}}& \multicolumn{4}{|l|}{\cellcolor[HTML]{BDD6EE} \textbf{\& Control }}& \multicolumn{3}{|l|}{\cellcolor[HTML]{BDD6EE}\textbf{Objectives }}   \\
 \multirow{15}{*}{\textbf{No.}}&
 \multirow{15}{*}{\textbf{Name of dataset}}&\multirow{15}{*}{\rotatebox[origin=c]{0}{\textbf{Year}}} &\multirow{8}{*}{\rotatebox[origin=c]{90}{\textbf{\# of attack types} }} & \multirow{4}{*}{\rotatebox[origin=c]{90}{TCP/UDP Port Scan }} & \multirow{4}{*}{\rotatebox[origin=c]{90}{Modbus / Scada Scan }} & \multirow{4}{*}{\rotatebox[origin=c]{90}{OS Fingerprint }} &\multirow{4}{*}{\rotatebox[origin=c]{90}{Vulnerability Scan }}  & \cellcolor[HTML]{FBE4D5}  &  \cellcolor[HTML]{FBE4D5}& \multirow{4}{*}{\rotatebox[origin=c]{90}{DoS /DDoS }}&\multirow{4}{*}{\rotatebox[origin=c]{90}{Injection-Protocol/Data}} &\multirow{4}{*}{\rotatebox[origin=c]{90}{Injection-SQL/XSS }} & \multirow{3}{*}{\rotatebox[origin=c]{90}{MITM }}&\multirow{4}{*}{\rotatebox[origin=c]{90}{PLC web service }} &\multirow{3}{*}{\rotatebox[origin=c]{90}{Replay }} &\multirow{3}{*}{\rotatebox[origin=c]{90}{DNP3 }} &\multirow{3}{*}{\rotatebox[origin=c]{90}{GOOSE }}&\multirow{3}{*}{\rotatebox[origin=c]{90}{S7 Comm }}&\multirow{4}{*}{\rotatebox[origin=c]{90}{SCADA/Modbus }} & \multirow{3}{*}{\rotatebox[origin=c]{90}{Backdoor }}&\multirow{3}{*}{\rotatebox[origin=c]{90}{Malware }}&\multirow{4}{*}{\rotatebox[origin=c]{90}{Brute Force }}&\multirow{4}{*}{\rotatebox[origin=c]{90}{Dictionary }} &\multirow{4}{*}{\rotatebox[origin=c]{90}{Malicious Insider }}&\multirow{3}{*}{\rotatebox[origin=c]{90}{Upload }}&\multirow{4}{*}{\rotatebox[origin=c]{90}{Exfiltration }}&\multirow{4}{*}{\rotatebox[origin=c]{90}{Tampering }}&\multirow{4}{*}{\rotatebox[origin=c]{90}{Ransomware }}\\
& & & & & &  & & \cellcolor[HTML]{FBE4D5}&\cellcolor[HTML]{FBE4D5} &  & & & & & & & & &  & & & &  & & & & & \\
& & & & & &  & & \cellcolor[HTML]{FBE4D5}&\cellcolor[HTML]{FBE4D5} &  & & & & & & & & &  & & & &  & & & & &\\
&  & & & & &  & & \cellcolor[HTML]{FBE4D5}&\cellcolor[HTML]{FBE4D5} &  & & & & & & & & &  & & & &  & & & & & \\
& & & & & &  & & \cellcolor[HTML]{FBE4D5}&\cellcolor[HTML]{FBE4D5}&  & & & & & & & & &  & & & &  & & & & &\\
& & & & & & & & \cellcolor[HTML]{FBE4D5}&\cellcolor[HTML]{FBE4D5}  &  & & & & & & & & &  & & & &  & & & & &\\
 && & & & & & & \cellcolor[HTML]{FBE4D5}&\cellcolor[HTML]{FBE4D5}  &  & & & & & & & & &  & & & &  & & & & &\\
& & & & & & & & \cellcolor[HTML]{FBE4D5}\multirow{-10}{*}{\rotatebox[origin=c]{90}{\textbf{Weaponisation} }}&\cellcolor[HTML]{FBE4D5}\multirow{-12}{*}{\rotatebox[origin=c]{90}{\textbf{Delivery} }} &  & & & & & & & & &  & & & &  & & & & &\\\hline
17 &\textbf{ICS Gas Pipeline}                       & 2013 & 11 &   & • &   &   & \cellcolor[HTML]{FBE4D5} & \cellcolor[HTML]{FBE4D5} & •         & • & \textbf{} & \textbf{} & \textbf{} & \textbf{} & \textbf{} & \textbf{} & \textbf{} & •                                            &   &   &   &   &   &   &   &   &                      \\
\rowcolor[HTML]{DEEAF6}18 & 
\textbf{ICS New Gas   Pipeline}                 & 2015 & 35 &   &   &   &   & \cellcolor[HTML]{FBE4D5} & \cellcolor[HTML]{FBE4D5} & •         & • & \textbf{} & •         & \textbf{} & \textbf{} & \textbf{} & \textbf{} & \textbf{} & •                                            &   &   &   &   &   &   &   &   &                      \\
19 &\textbf{ICS Power System}                       & 2014 & 3  &   &   &   &   & \cellcolor[HTML]{FBE4D5} & \cellcolor[HTML]{FBE4D5} & \textbf{} & • & \textbf{} & \textbf{} & \textbf{} & \textbf{} & \textbf{} & \textbf{} & \textbf{} & •                                            &   &   &   &   &   &   &   &   &                      \\
\rowcolor[HTML]{DEEAF6} 20 &
\textbf{IEC 60870-5-104}                        & 2020 & 14 &   &   &   &   & \cellcolor[HTML]{FBE4D5} & \cellcolor[HTML]{FBE4D5} & •         & • &           & •         &           &           &           &           &           &                                              &   &   &   &   &   &   &   &   &                      \\
21 &\textbf{IEC 61850 Security}                     & 2019 & 6  &   &   &   &   & \cellcolor[HTML]{FBE4D5} & \cellcolor[HTML]{FBE4D5} & •         & • &           &           &           & •         &           & •         &           &                                              &   &   &   &   &   &   &   &   &                      \\
\rowcolor[HTML]{DEEAF6}22 & 
\textbf{ISOT}                                        & 2022 & 6  &   &   &   &   & \cellcolor[HTML]{FBE4D5} & \cellcolor[HTML]{FBE4D5}  & •         & • &           & •         &           &           &           &           &           &                                              &   &   &   &   &   &   &   &   &                      \\
23 &\textbf{Lemay Modbus}                           & 2016 & 5   &   & • &   &   & \cellcolor[HTML]{FBE4D5} & \cellcolor[HTML]{FBE4D5} &            & • &           & •         &           &           &           &           &          & •                                            &   & • &   &   &   &   &   &   &   \\
\rowcolor[HTML]{DEEAF6}24 & 
\textbf{Maynard SCADA}                          & 2018 & 5  & • & • &   &   & \cellcolor[HTML]{FBE4D5} & \cellcolor[HTML]{FBE4D5} &           & • &           & •         &           &           &           &           &           & •                                            &   &   &   &   &   &   &   &   &                      \\
25 &\textbf{Modbus TCP SCADA \#1}                   & 2018 & 4  &   &   &   &   & \cellcolor[HTML]{FBE4D5} & \cellcolor[HTML]{FBE4D5} & •         &   &           & •         &           &           &           &           &           & •                                            &   &   &   &   &   &   &   &   &                      \\
\rowcolor[HTML]{DEEAF6} 26&
\textbf{Realtime ICS SCADA}                     & 2021 & 4  & • &   & • &   & \cellcolor[HTML]{FBE4D5} & \cellcolor[HTML]{FBE4D5} &           &   &           & •         & •         &           & •         &           &           & •                                            &   & • &   &   &   &   &   &   &                      \\
27 &\textbf{SWaT}                                   & 2017 & 41 &   &   &   &   & \cellcolor[HTML]{FBE4D5} & \cellcolor[HTML]{FBE4D5} &           & • &           & •         &           &           &           &           &           & •                                            &   &   &   &   &   &   &   &   &                      \\
\rowcolor[HTML]{DEEAF6} 28 &
\textbf{TON\_IoT}                               & 2020 & 2  & • & • &   &   & \cellcolor[HTML]{FBE4D5} & \cellcolor[HTML]{FBE4D5} & •         &   & •         &           &           &           &           &           &           & •                                            & • &   & • & • &   &   &   & • &                      \\
29 &\textbf{WADI}                                   & 2017 & 15 &   &   &   &   & \cellcolor[HTML]{FBE4D5} & \cellcolor[HTML]{FBE4D5} & •         & • &           & •         &           &           &           &           &           & •                                            &   &   &   &   &   &   &   &   &                      \\
\rowcolor[HTML]{DEEAF6}30 &
\textbf{WUSTL-IIOT-2018   ICS (SCADA)}          & 2020 & 5  & • & • & • &   & \cellcolor[HTML]{FBE4D5} & \cellcolor[HTML]{FBE4D5} &           & • &           &           &           &           &           &           &           & •                                            &   &   &   &   &   &   &   &   &                      \\
31 &\textbf{WUSTL-IIOT-2021}                        & 2021 & 4  & • &   &   &   & \cellcolor[HTML]{FBE4D5} & \cellcolor[HTML]{FBE4D5} & •         & • &           &           &           &           &           &           &           &                                              & • &   &   &   &   &   &   &   &                      \\
\rowcolor[HTML]{DEEAF6} 32 &
\textbf{X-IIOTID}                               & 2020 & 19 & • &   & • & • & \cellcolor[HTML]{FBE4D5} & \cellcolor[HTML]{FBE4D5} & •         &   &           & •         &           &           &           &           &           & •                                            &   &   & • & • & • & • & • & • & •                      \\\hline              

\end{tabular}
 
}

\end{table}
\end{landscape}